\documentclass[showpacs,aps,twocolumn]{revtex4}
\usepackage{epsfig}
\usepackage{graphicx}
\usepackage{amsmath,amssymb,amsfonts}
\usepackage{array}
\usepackage{url}
\usepackage{hyperref}
\usepackage{multirow}
\usepackage{float}
\usepackage{lineno}
\usepackage{xspace}
\usepackage[usenames,dvipsnames]{color}

\newcommand{\bef}{\begin{figure}}
\newcommand{\eef}{\end{figure}}
\newcommand{\bc}{\begin{center}}
\newcommand{\ec}{\end{center}}

\newcommand{\be}{\begin{equation}}
\newcommand{\ee}{\end{equation}}
\newcommand{\bea}{\begin{eqnarray}}
\newcommand{\eea}{\end{eqnarray}}

\begin{document}
\title{Investigating Heavy-flavor vs Light-flavor Puzzle with Event Topology and Multiplicity in Proton+Proton Collisions at $\sqrt{s}$ = 13 TeV using PYTHIA8}
\author{Suman Deb}
\author{Raghunath Sahoo\footnote{Presently CERN Scientific Associate, CERN, Geneva, Switzerland}}
\email{Corresponding author\\ Email: Raghunath.Sahoo@cern.ch}
\affiliation{Department of Physics, Indian Institute of Technology Indore, Simrol, Indore 453552, INDIA}
\author{Dhananjaya Thakur}
\affiliation{Institute of Modern Physics, Chinese Academy of Sciences, Lanzhou, China}
\author{Sushanta Tripathy\footnote{Presently at INFN - sezione di Bologna, via Irnerio 46, 40126 Bologna BO, Italy}}
\affiliation{Instituto de Ciencias Nucleares, UNAM, Deleg. Coyoac\'{a}n, Ciudad de M\'{e}xico 04510}
\author{Arvind Khuntia}
\affiliation{The H. Niewodniczanski Institute of Nuclear Physics, Polish Academy of Sciences, PL-31342 Krakow, Poland}

\begin{abstract}
Heavy-flavored hadrons are unique probes to study the properties of hot and dense QCD medium produced in ultra-relativistic heavy-ion collisions at RHIC and the LHC. Transverse spherocity is one of the event-topology variables used to separate jetty and isotropic events from the pool of event samples. This study aims to understand the production dynamics of heavy-flavors  through the transverse momentum spectra, double differential yield and mean transverse momentum of J/$\psi$, $\rm D^{0}$ and $\Lambda_{c}^{+}$ as a function of charged-particle multiplicity and transverse spherocity. Further to investigate the possibility of hadronization of the charm quarks,  transverse spherocity dependence ratios like $\Lambda_{c}^{+}$/$\rm D^{0}$ and $\Lambda^{0}$/$K^{-}$ are studied. For the current analysis, the events are generated by using 4C tuned PYTHIA8 for pp at $\sqrt{s}$ = 13 TeV, which is quite successful in explaining the heavy-flavor particle production at the LHC energies.
 \pacs{25.75.Dw,14.40.Pq}
\end{abstract}
\date{\today}
\maketitle

\section{Introduction}
\label{intro}
Heavy-flavor hadrons,  containing  open  or  hidden  charm and beauty flavors are believed to be important probes for the  understanding of  Quantum Chromodynamics (QCD)  in  high-energy hadronic collisions: starting from the study of production mechanisms in  proton-proton (pp) collisions to the investigation of Cold Nuclear Matter (CNM) effects in proton-nucleus (p--A) collisions and their suppression in the search of Quark Gluon Plasma (QGP) in nucleus-nucleus (A--A) collisions~\cite{Xu:2017hgt,Adare:2012yxa,Tang:2020ame}. In addition, the study of heavy-flavor production as a function of the charged-particle multiplicity may provide insights into multiple hard partonic scatterings~\cite{Abelev:2012rz,Acharya:2020pit,Adam:2015ota}. Recently, the observation of heavy-ion-like features in small systems (pp and p$-$A) continues to generate considerable interest in the scientific community. For example, the discovery of collective-like phenomena~\cite{Li:2011mp,Khachatryan:2010gv}, strangeness enhancement~\cite{ALICE:2017jyt} etc., and corresponding  phenomenological studies~\cite{Deb:2020ezw,Mishra:2020epq} in high-multiplicity pp and p$-$A collisions are few among them. In this context, the observed QGP-like phenomena warrants a deeper understanding involving many complex dynamical processes like resonance decays, jets, underlying events (UE) etc. Therefore, small systems need to be re-investigated properly including the light and heavy-flavor sectors, as the production dynamics of these sectors are different in nature. To observe similar effects and in particular, the interplay of hard processes and UE, heavy-flavors are very useful tools. The study of heavy-flavor transverse momentum ($p_{\rm T}$) spectra is one of the main tools to disentangle collective effects from trivial correlations. The production dynamics of heavy-flavor can be better understood through a differential analysis involving tools to separate jetty and isotropic events~\cite{Cuautle:2015kra}. Phenomenological Monte Carlo generators such as PYTHIA8~\cite{Pythia} are widely used to compare expectations where different known physics mechanisms are taken into account. Multi-Parton Interactions (MPIs) together with Color Reconnection (CR) in PYTHIA8 could mimic collective-like expansion~\cite{Ortiz:2013yxa}. In this present paper, we have performed a multi-differential study of heavy-flavor production (J/$\psi$, $\rm D^{0}$ and $\Lambda_{c}^{+}$) as a function of charged-particle multiplicity and transverse spherocity using pQCD-inspired event generator, PYTHIA8~\cite{Pythia}.

As J/$\psi$ suppression is a potential probe of QGP \cite{Matsui:1986dk}, leading to enhancement of open charms like $\rm D^{0}$ and $\Lambda_{c}^{+}$, this
multi-differential study in pp collisions taking minimum-bias (MB) and high-multiplicity events will be of real help in understanding the underlying
production dynamics of heavy-flavors.



The  simulation of pp collisions implemented in PYTHIA8 starts with the basic concept, p+p $\rightarrow$ n, defining ``n" particles produced by the collision of two protons.  If we consider a proton, then in first approximation one would get distribution of quarks: u, u, and d. But in reality, protons are active soup of partons: quark-antiquark pairs appearing from vacuum or gluons, which create so-called sea of particles inside a proton. Larger fraction of low-$x$ gluons and sea quarks start to appear in case of higher $Q^{2}$ (higher colliding  energy) coming from a parton splitting. Multi-partonic interactions (MPI) is the natural consequence of this composite nature of proton~\cite{Sjostrand:2007gs}. MPIs are mainly responsible for soft underlying events, but there is finite probability of occurrence of several hard scattering processes in the same hadron-hadron collisions~\cite{Sjostrand:2007gs,Butterworth:1996pi,Adam:2015ota}. The large number of MPIs from pp collisions generate a very high density of QCD matter. This concept is different from the traditional one, which has back-to-back jet production in pp collisions. Therefore, it is worth separating these MPI dominant events (isotropic) from jet-rich (jetty) events and revisit the pp results at the LHC. Surprisingly, current PYTHIA8 event topology studies have shown that high-multiplicity pp collisions exhibit prominent features of heavy-ion collisions~\cite{Ortiz:2017jho,Acharya:2019mzb,Tripathy:2020jue}. This questions the interpretation of heavy-ion results which relies heavily on the results obtained in minimum bias pp collisions. The Run1 and Run2 of the LHC have provided a large number of measurements to explore the pp physics, probing both soft and hard QCD regime, and thereby a few tests on possible QGP signatures in pp collisions. There are still some phenomena like baryon-to-meson ratios, degree of collectivity etc., that are yet to be understood completely. It is now an ideal time to perform these studies further using realistic Monte Carlo event generators and open windows for such measurements for the future LHC Run3. One of such interesting puzzle is an enhanced production of heavy baryons over mesons ratio in the intermediate-$p_{\rm{T}}$ region in central heavy-ion collisions at the midrapidity~\cite{Acharya:2017kfy,Adam:2019hpq,Oh:2009zj}. This enhancement is described by multi-quark dynamics through quark coalescence or recombination~\cite{Greco:2003mm,Oh:2009zj}. Minijet partons produced in relativistic heavy-ion collisions coalesce with partons from the QGP expected to be formed in relativistic heavy-ion collisions leading to an enhanced production of hadrons with intermediate transverse momentum~\cite{Greco:2003xt}. Therefore, the enhanced production of baryon over meson is thought to be an indication of QGP formation in heavy-ion collisions. Recently, LHC experiments have also observed the similar signature for charmed baryon-to-meson ratio for Pb-Pb and p-Pb collisions~\cite{Acharya:2018ckj,Aaij:2018iyy,Acharya:2017kfy}. Such ratio is also measured in pp collisions~\cite{Acharya:2020lrg}. Taking the opportunity to use transverse spherocity, a tool which separates jet-rich events from isotropic events (rich in MPI), we have analysed the charmed baryon-to-meson ratio in pp collisions using PYTHIA8 and looked for existence of such enhancement in intermediate-$p_{\rm{T}}$ region. Further, we have analysed its counterpart ratio in light flavor sector to know the behaviour and hence the production mechanism when we go from light flavor to heavy-flavor sector. 

The paper is organised as follows. We begin with a brief motivation for the study in Section~\ref{intro}. In Section~\ref{eventgen}, the detailed analysis methodology along with brief description about PYTHIA8 are given. Section~\ref{result} discusses about the results and finally they are summarized in Section~\ref{sum}. In the appendix, we have compared the production cross-sections of J/$\psi$ and $\rm D^{0}$ between experimental data from ALICE and PYTHIA8 in the same kinematic range to have a glimpse of the spectral shapes, which are essential for a $p_{\rm T}$-differential study.

\section{Event generation and Analysis methodology}
\label{eventgen}
PYTHIA8, a standalone Monte-Carlo event generator, is widely used to simulate ultra-relativistic collisions among the particles like electron-electron, electron-positron, proton-proton and proton-antiproton. It has been quite successful in explaining many of the experimental data from the LHC qualitatively with different incorporated physics processes. 
PYTHIA8 includes Multi-Parton Interaction (MPI) scenario, which allows heavy-flavor quarks to be produced through $2\rightarrow2$ hard 
sub-processes. Detailed explanation on PYTHIA8 physics processes and their implementation can be found in Ref.~\cite{Pythia}. 

The results reported in this paper are obtained from simulated inelastic, non-diffractive events using PYTHIA version 8.215~\cite{Sjostrand:2014zea} with the 4C tune (Tune:pp = 5)~\cite{Corke:2010yf}. Further, non-diffractive component of the total cross section for all hard QCD processes (HardQCD:all=on) are considered, which includes the production of heavy quarks along with MPI-based scheme of color reconnection (ColourReconnection:reconnect = on). A cut of $p_{\rm T}\geq$ 0.5 GeV/$c$ (using PhaseSpace:pTHatMinDiverge) is used to avoid the divergences of QCD processes in the limit $p_{\rm{T}}$ $\rightarrow$ 0. For the production of quarkonia through NRQCD framework~\cite{Shao:2012iz,Caswell:1985ui,Bodwin:1994jh}, we use Charmonium:all flag in the simulation. Study of $\rm D^{0}$, J/$\psi$ and $\Lambda_{c}^{+}$ production are done at the mid-rapidity. J/$\psi$, $\rm D^{0}$ and $\Lambda_{c}^{+}$ are reconstructed via the $e^{+}+e^{-}$ ($|y|<$ 0.9)~\cite{Lofnes}, $K^{-}+\pi^{+}$ ($|y|<$ 0.5)~\cite{Hamon:2018zqs} and $p+K^{-}+\pi^{+}$ ($|y|<$ 0.5)~\cite{Acharya:2017kfy} decay channels and their yields are obtained through invariant mass reconstruction keeping the detector acceptance of ALICE in mind. This analysis is performed by generating 100 million events for J/$\psi$ and approximately 50 million events each for $\rm D^{0}$ and $\Lambda_{c}^{+}$ in pp collisions at $\sqrt{s}$ = 13 TeV.  The charged-particle multiplicity, $N_{\rm ch}$ is measured at the mid-rapidity ($|\eta|<$ 1.0), considering all the charged particles including the decay product of J/$\psi$, $\rm D^{0}$ and $\Lambda_{c}^{+}$. As the aim of this work is to perform a multi-differential study using transverse spherocity and the charged-particle multiplicities ($N_{\rm ch}$), we have chosen the minimum bias (0-100\%) collisions and events with top $20\%$ of $N_{\rm ch}$ for our study.

Transverse spherocity ($S_{0}$) is defined for a unit vector $\hat{n} (n_{T},0)$ that minimizes the following ratio~\cite{Cuautle:2015kra,Salam:2009jx}.
\begin{eqnarray}
S_{0} = \frac{\pi^{2}}{4} \bigg(\frac{\Sigma_{i}~|\vec p_{\rm T_{i}}\times\hat{n}|}{\Sigma_{i}~p_{\rm T_{i}}}\bigg)^{2}.
\label{eq1}
\end{eqnarray}

\begin{figure}[ht!]
\includegraphics[scale=0.43]{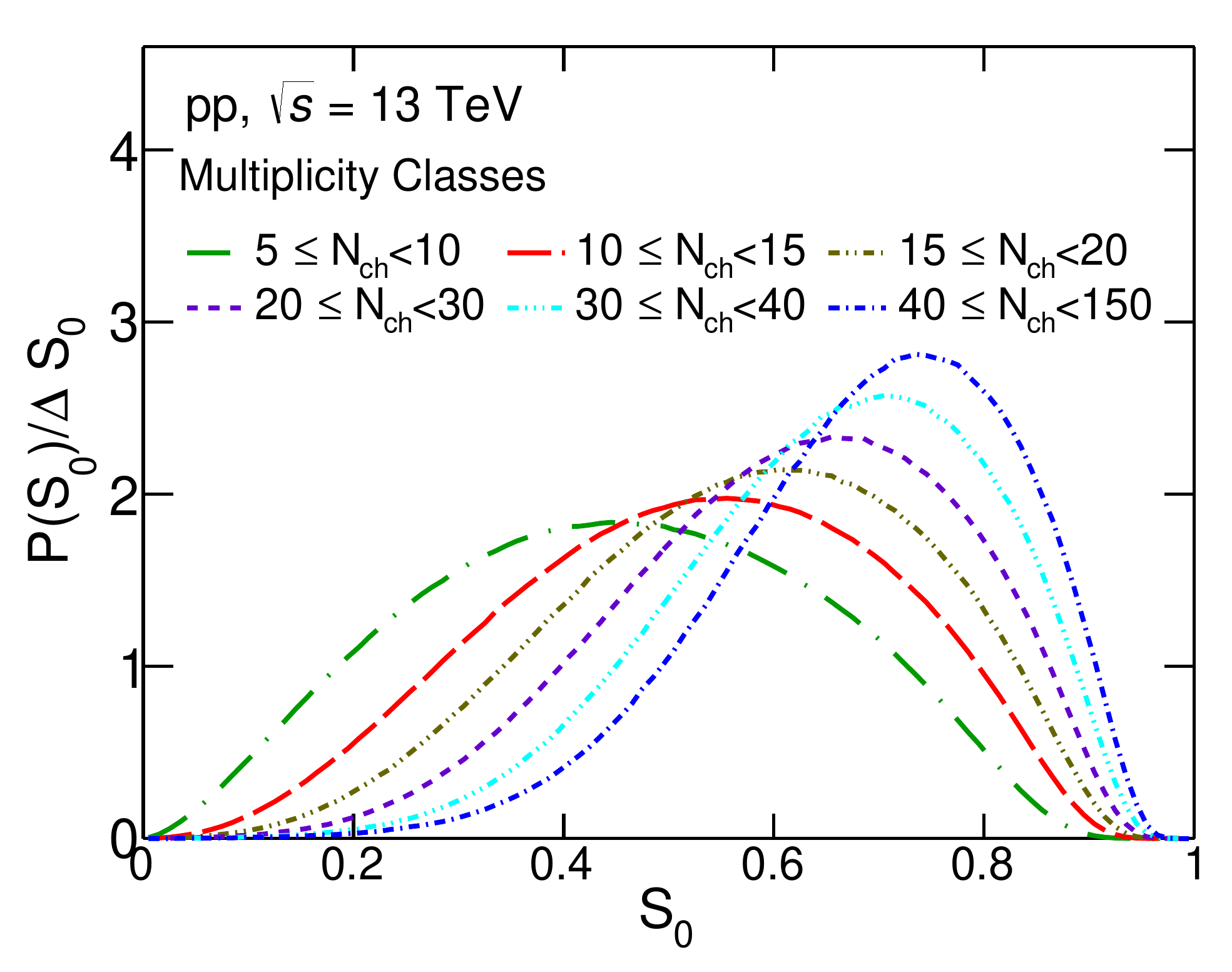}
\caption[]{(Color Online) Transverse spherocity distributions for different charged-particle multiplicity in pp collisions at $ \sqrt{s} =\mathrm{13~TeV}$ using PYTHIA8. Different line styles and colors are for different multiplicity classes. }
\label{sp_dis}
\end{figure}

By construction, $S_{0}$ is infrared and collinear safe~\cite{Salam:2009jx} and it ranges from 0 to 1. $S_{0}$ becoming 0 are referred to the jetty events while 1 would mean the events are isotropic in nature. As transverse spherocity distinguishes events based on different topological limits $i.e.$ events with back-to-back jet structures (jetty) versus events dominated by multiple soft scatterings (isotropic), in this analysis only the events with at least 5 charged-particles in $|\eta|<$ 0.8 with $p_{\rm T}>$ 0.15 GeV/$c$ are considered, so that the concept of event topology becomes statistically meaningful. $S_{0}$ cuts on the generated events are applied in order to sort out jetty and isotropic events from the total events. For minimum bias collisions, the cuts for jetty events is $0 \leq S_{0} < 0.37$ with lowest 20\% of $S_{0}$ distribution and $0.72 < S_{0} \leq 1$ is for isotropic events with highest 20\% of $S_{0}$ distribution. Further, minimum bias events are divided into six multiplicity classes and the corresponding spherocity cuts for isotropic and jetty events are tabulated in Table~\ref{tab:mult_sp}. For consistency, $N_{\rm ch}$ intervals chosen here are the same as in Ref.~\cite{Khatun:2019dml}. In order to maximize the statistics, the bin-width is taken smaller at lower multiplicities and then subsequently higher at high multiplicity bins. Figure~\ref{sp_dis} represents the transverse spherocity distribution in different multiplicity classes for pp collisions at  $\sqrt{s}$ = 13 TeV. Here, it is observed that high-multiplicity events are more towards isotropic in nature which is in accordance with earlier works on transverse spherocity~\cite{Khatun:2019dml,Khuntia:2018qox,Tripathy:2019blo}. The peak of the transverse spherocity distribution shifts towards isotropic events with increasing charged-particle multiplicity. This shows that higher contribution of softer events come from multiple hard partonic scatterings in high-multiplicity pp collisions, which generate an almost isotropic distribution of particles~\cite{Khuntia:2018qox}. Therefore, the differential study of particle production as a function of multiplicity and event shape classes has great importance to understand the particle production mechanism. As transverse spherocity distribution depends on charged-particle multiplicity, the cuts for jetty and isotropic events vary for different transverse spherocity classes, which is shown in Table~\ref{tab:mult_sp}. For the sake of simplicity, here onwards we refer transverse spherocity as spherocity.

\begin{table}[htp]
\caption{Charged-particle multiplicity (Mult.) classes ($N_{\rm ch}$) ($|\eta|<$ 1.0) and corresponding spherocity ranges for jetty and isotropic events. Here the lowest and highest 20\% events of the spherocity distribution for a given multiplicity class are considered as jetty and isotropic events, respectively.}
\centering 
\scalebox{1.17}
{
\begin{tabular}{|c|c|c|}
\hline  
\multicolumn{1}{|c|}{\bf Mult. Classes }&\multicolumn{2}{c|}{\bf $S_{0}$ range} \\
\cline{2-3}
\multicolumn{1}{|c|}{($N_{\rm ch}$)} &{\bf Jetty events} & {\bf Isotropic events}\\
\hline

\multirow{1}{*}{$5-10$}  

&$0-0.29$ &$0.64-1$ \\
\cline{2-3} 
\cline{1-3} 

\multirow{1}{*}{$10-15$} 

& $0-0.38$ &$0.70-1$ \\
\cline{2-3} 
\cline{1-3} 

\multirow{1}{*}{$15-20$} 

&$0- 0.44$&$0.74-1$\\
\cline{2-3} 
\cline{1-3} 

 \multirow{1}{*}{$20-30$} 

&$0-0.49$ &$0.77- 1$ \\

\cline{2-3} 
\cline{1-3} 

 \multirow{1}{*}{$30-40$} 

&$0- 0.54$&$0.80- 1$ \\
\cline{2-3} 
\cline{1-3} 

 \multirow{1}{*}{$40-150$} 

&$0- 0.58$ &$0.82- 1$ \\
\cline{2-3} 
\cline{1-3} 

 \end{tabular}
}
\label{tab:mult_sp}
\end{table}

With this detailed analysis methodology, we now proceed for the estimation of transverse momentum spectra, relative integrated yield and relative mean transverse momentum of $\rm D^{0}$, J/$\psi$ and $\Lambda_{c}^{+}$ in pp collisions at $\sqrt{s}$ = 13 TeV.

 \section{Results and Discussion}
 \label{result}

 \subsection{Transverse momentum spectra}
  \label{hard_qcd}
  
  \begin{figure*}[!ht]
\bc
  \includegraphics[scale=0.66]{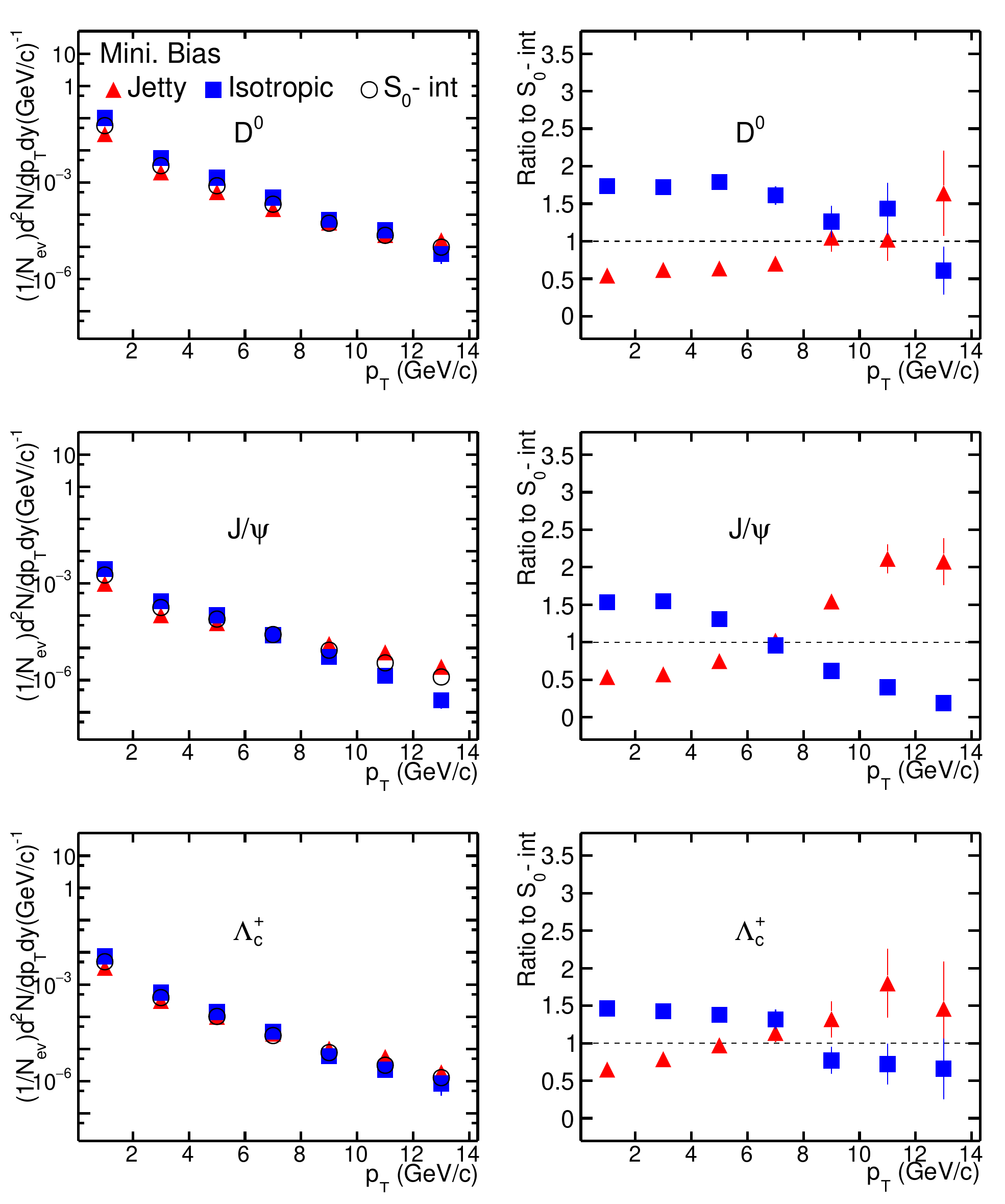}
 \caption  {(Color online) $p_{\rm T}$-spectra (left panel) of isotropic, jetty and spherocity-integrated events, and their ratios (right panel) to the spherocity-integrated ones for $\rm D^{0}$ (top), J/$\psi$ (middle) and $\Lambda_{c}^{+}$ (bottom) for minimum bias pp collisions at $\sqrt{s}$ = 13 TeV using PYTHIA8. }
 \label{fig:pTmin}  
 \ec
 \end{figure*}
 
   \begin{figure*}[!ht]
\bc
  \includegraphics[scale=0.66]{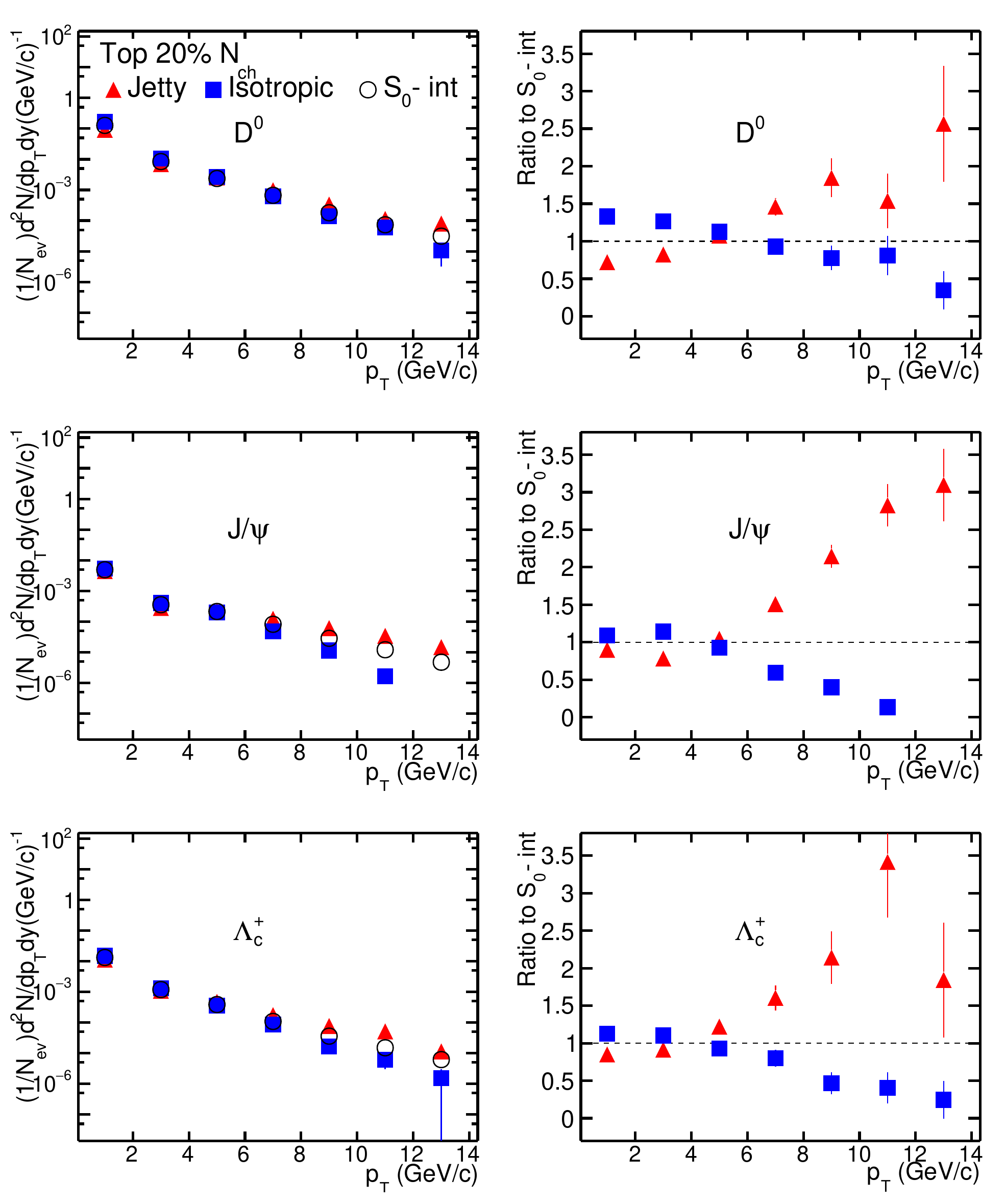}
 \caption  {(Color online) $p_{\rm T}$-spectra (left panel) of isotropic, jetty and spherocity-integrated events, and their ratios  (right panel) to the spherocity-integrated ones for $\rm D^{0}$ (top), J/$\psi$ (middle) and $\Lambda_{c}^{+}$ (bottom) for high-multiplicity pp collisions at $\sqrt{s}$ = 13 TeV using PYTHIA8. }
 \label{fig:pTtop20}  
 \ec
 \end{figure*}
 
 Left panels of Fig. \ref{fig:pTmin} show the transverse momentum ($p_{\rm T}$) spectra for $\rm D^{0}$ (top), J/$\psi$ (middle) and $\Lambda_{c}^{+}$ (bottom) for isotropic, jetty and spherocity-integrated events in minimum bias pp collisions at $\sqrt{s}$ = 13 TeV. The right panels show the ratio of the $p_{\rm T}$ spectra for isotropic and jetty events to the spherocity-integrated ones. The ratios clearly indicate that the particle production from isotropic events dominate at low-$p_{\rm T}$ and after a certain $p_{\rm T}$, the particle production from jetty events starts to dominate. The crossing point of the jetty and isotropic events for $\Lambda_{c}^{+}$ and J/$\psi$ are found to be similar. However, for $\rm D^{0}$ the crossing point is at a higher $p_{\rm T}$. This may suggest that the soft production of $\rm D^{0}$ is dominant till higher-$p_{\rm T}$ compared to $\Lambda_{c}^{+}$ and J/$\psi$. We also estimate the $p_{\rm T}$ spectra in high-multiplicity pp collisions in different spherocity classes, which is shown in Fig. \ref{fig:pTtop20}. Here, the crossing point of the jetty and isotropic events for all the studied particles are found to be similar. The comparison between Fig. \ref{fig:pTmin} and Fig. \ref{fig:pTtop20} indicates that the heavy-flavor particle production from jetty events dominates at a lower $p_{\rm T}$ in high-multiplicity pp collisions compared to the minimum bias ones. At high multiplicity for low-$p_{\rm T}$ region, the separation between the isotropic and jetty events are small compared to minimum bias events. 

 \subsection{Relative integrated yield and relative mean transverse momentum}
 \label{yield_pt_mult}

\begin{figure*}[!ht]
\bc
\includegraphics[scale=0.46]{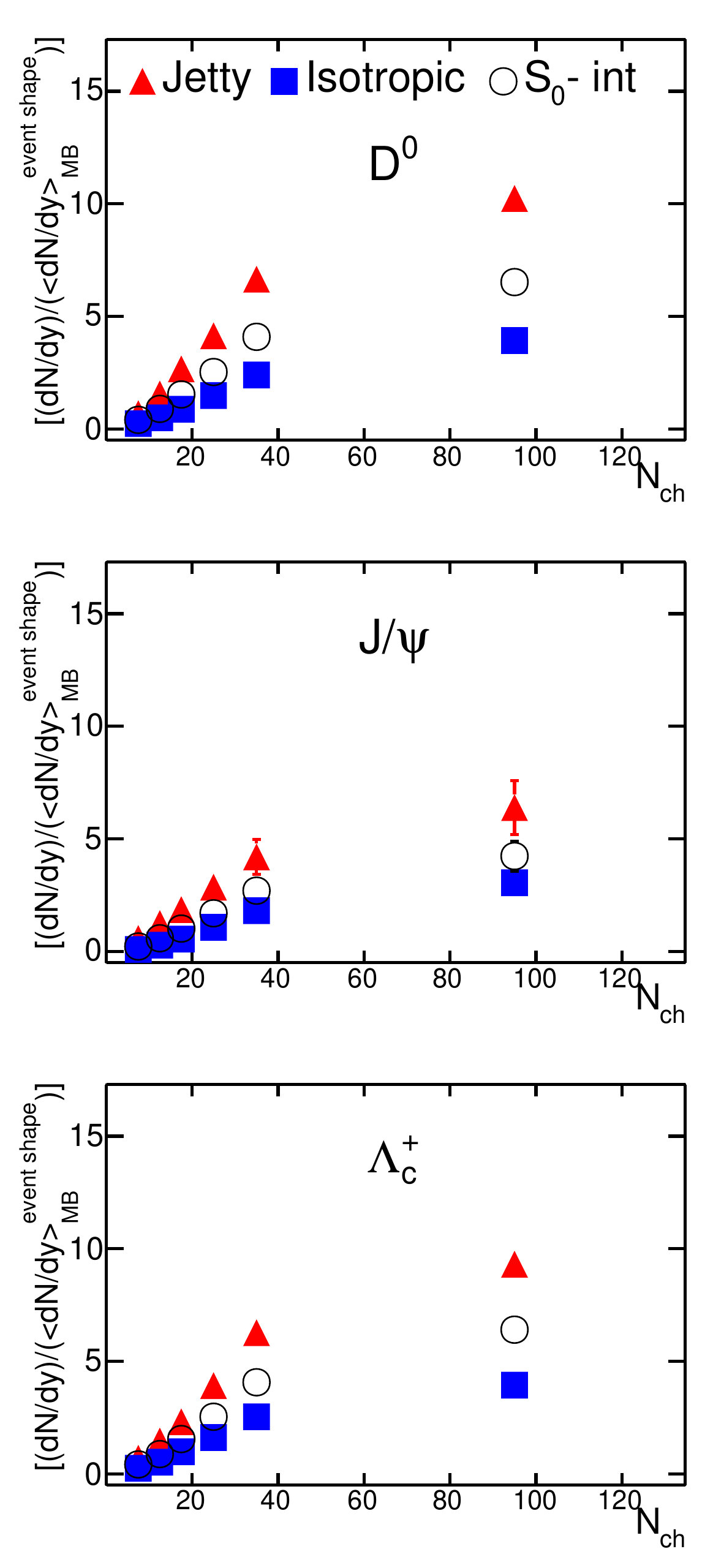}
\includegraphics[scale=0.66]{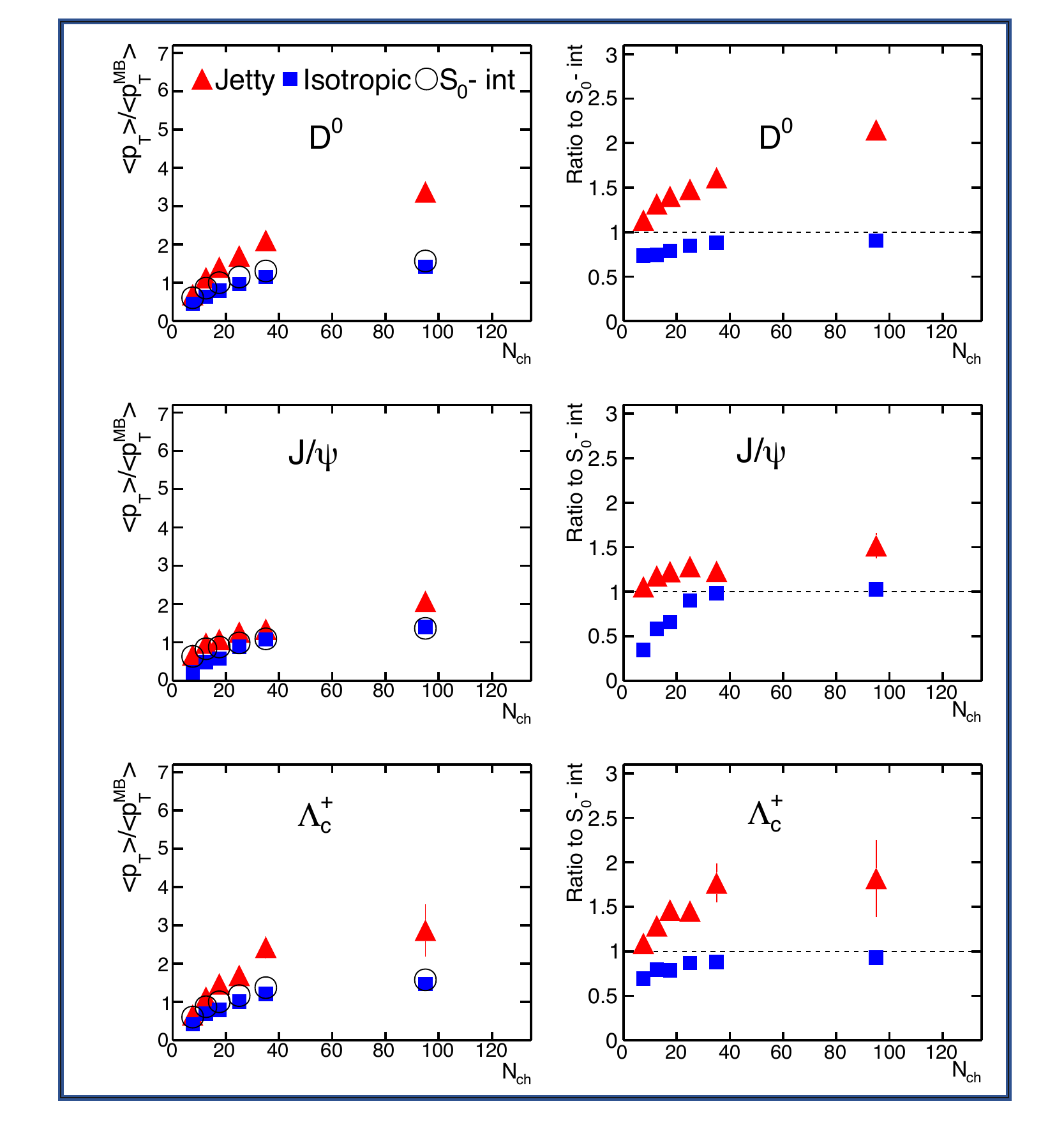}
\caption {(Color online) Left panel: Self-normalised yields with respect to the corresponding event types, Middle panel: mean transverse momenta ($\langle p_{\rm{T}} \rangle$) scaled to its MB values, and Right panel: ratio of $\langle p_{\rm{T}} \rangle$ in different event types to the spherocity-integrated ones as a function of multiplicity for $\rm D^{0}$ (top), J/$\psi$ (middle) and $\Lambda_{c}^{+}$ (bottom). The error bars in the data points are the statistical uncertainties.}
\label{fig5:meanpt:spherocity}  
\ec
\end{figure*}    

The relative yields of $\rm D^{0}$, J/$\psi$ and $\Lambda_{c}^{+}$ are measured at the mid-rapidity ($|y| < 0.9$) using the following relation:
\bea
\frac{Y_{particle}}{<Y_{particle}>}= \frac{N_{particle}^{i}}{N_{particle}^{total}}\frac{N_{evt}^{total}}{N_{evt}^{i}},
\label{eq1}
\eea
where, $N_{\rm particle}^{i}$ and $N_{\rm evt}^{i}$ are the number of $\rm D^{0}$, J/$\psi$ and $\Lambda_{c}^{+}$, and number of events in $i^{th}$ multiplicity bin, respectively. $N_{\rm particle}^{\rm total}$ and $N_{\rm evt}^{\rm total}$ are the total number of $\rm D^{0}$, J/$\psi$ and $\Lambda_{c}^{+}$ produced, and total number of minimum-bias events, respectively. The uncertainties in the measurement of the number of $\rm D^{0}$, J/$\psi$ and $\Lambda_{c}^{+}$ particles are $\sqrt{N_{\rm D^{0}}}$, $\sqrt{N_{J/\psi}}$ and $\sqrt{N_{\Lambda_{c}^{+}}}$, respectively . These uncertainties are propagated using standard error propagation formula to estimate the uncertainties in relative $\rm D^{0}$, J/$\psi$ and $\Lambda_{c}^{+}$ yields. The mean transverse momenta ($\langle p_{\rm{T}}\rangle$) of $\rm D^{0}$, J/$\psi$ and $\Lambda_{c}^{+}$ are calculated for each multiplicity bin and corresponding uncertainty is given by the ratio of standard deviation ($\sigma$) and square root of the number of entries in that bin ($\sigma/\sqrt{N_{\rm bin}^{p_{\rm T}}}$).

Left(middle) panel of Fig. \ref{fig5:meanpt:spherocity} shows the integrated yields($\langle p_{\rm{T}}\rangle$) of $\rm D^{0}$, J/$\psi$ and $\Lambda_{c}^{+}$ scaled to the corresponding integrated yields($\langle p_{\rm{T}}\rangle$) of spherocity-integrated events in minimum bias collisions as a function of charged-particle multiplicity. For all the particles, the relative yield and the relative mean transverse momentum increase with charged-particle multiplicity. Enabling the CR in PYTHIA8, produces effects on the final particle distributions, which could resemble those due to flow~\cite{Ortiz:2013yxa}. An increase in the $\langle p_{\rm{T}}\rangle$ with  $N_{\rm ch}$ is attributed to the presence of CR between the interacting strings.  The relative yields and relative $\langle p_{\rm{T}}\rangle$ are found to be higher for jetty events compared to isotropic ones. The  right panel of Fig. \ref{fig5:meanpt:spherocity} shows the ratio of relative mean transverse momentum from isotropic and jetty events to the $S_{0}$-integrated events. Interestingly, the relative $\langle p_{\rm{T}}\rangle$ of the studied particles for isotropic events stay systematically below the spherocity-integrated ones for low-multiplicity events and approaches towards spherocity integrated ones with increase of multiplicity. For jet-like events the $\langle p_{\rm{T}}\rangle$ is higher than that of spherocity-integrated events and the relative increase in $\langle p_{\rm{T}}\rangle$ saturates at high multiplicity. This behavior is similar to the observed behavior of $\langle p_{\rm{T}}\rangle$ of light-flavor 
charged-particles in different spherocity classes by ALICE at the LHC~\cite{Acharya:2019mzb}.

Figure~\ref{fig5:meanpt:spherocity} further reveals a clear distinction in the production mechanisms between $\rm D^{0}$ and $\Lambda_{c}^{+}$ versus J/$\psi$. For example, relative yields of J/$\psi$ for jetty, isotropic and $S_{0}$-integrated events are close to each other and are less than that of $\rm D^{0}$ and $\Lambda_{c}^{+}$. This means more number of open flavors are produced in high-multiplicity events as compared to charmonia and is also reflected in the complementary study of $\langle p_{\rm{T}}\rangle$. The $\langle p_{\rm{T}}\rangle$ of J/$\psi$ has the dominant effect of jetty events, whereas, $\langle p_{\rm{T}}\rangle$ of $\rm D^{0}$ and $\Lambda_{c}^{+}$ are dominated by isotropic ones. This can be explained by the multi-quark dynamics by the fact that  $\rm D^{0}$-mesons and $\Lambda_{c}^{+}$ baryons are produced via string fragmentation. Here, the latter carry the flow-like characteristics originating from CR mechanism~\cite{Ortiz:2013yxa}. But, J/$\psi$ which is a bound state of heavy charm and anti-charm quarks, has a very little contribution from CR~\cite{Thakur:2017kpv,Deb:2018qsl}. Further, greater number of light-quarks are produced from MPI compared to heavy-quarks, which makes more light quarks to come to the close proximity of a c-quark, as compared to its own counter part ($\bar{c}$) and hence higher probability of production of open heavy-flavors than charmonia in a high-multiplicity environment. However, enhancement of heavy-baryon over meson still need to be understood which we have tried to explore in the next section.

\subsection{Baryon-to-meson ratio}
\label{sec:yield_ratio}
\begin{figure}[!ht]
\bc
\includegraphics[scale=0.438]{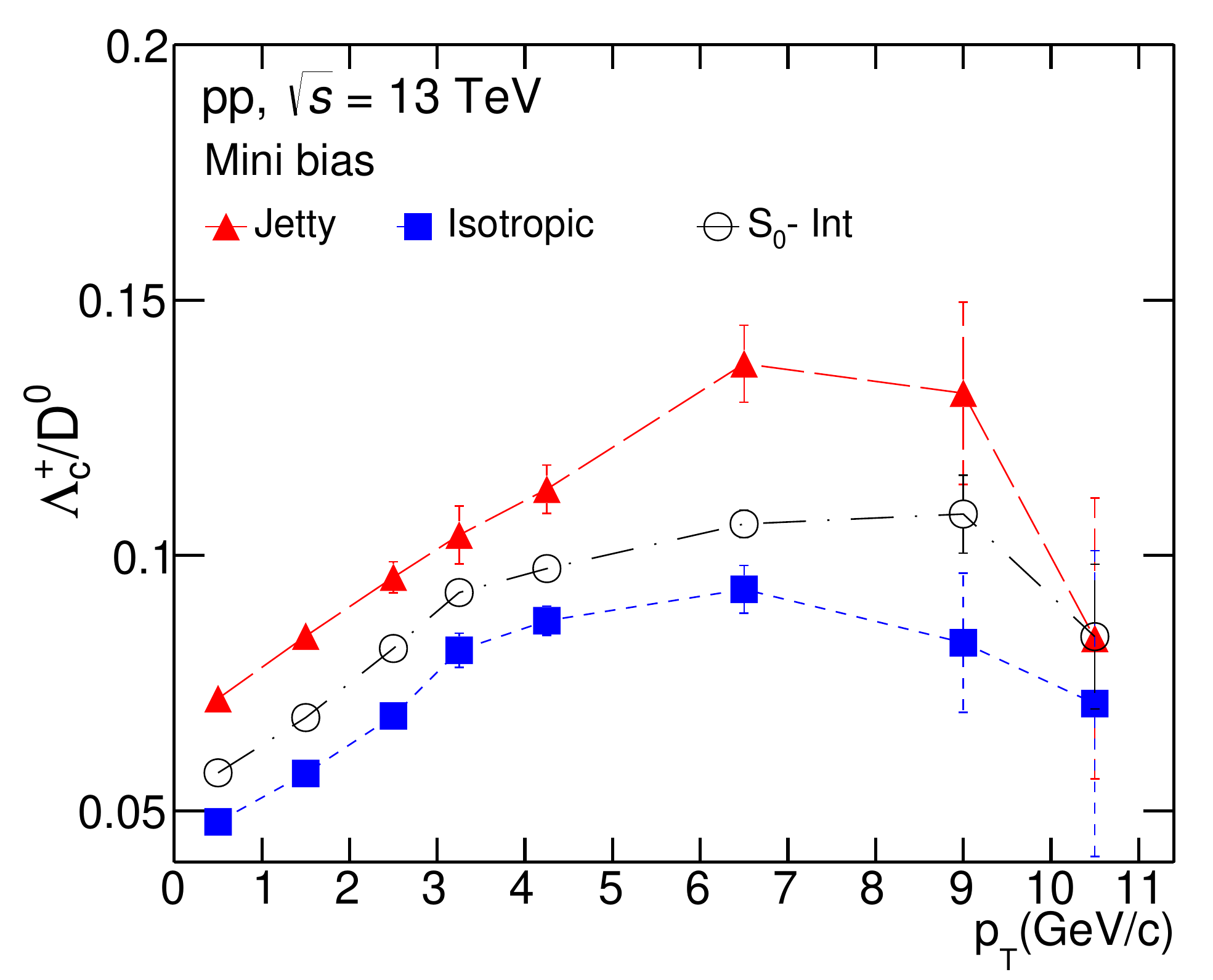}
\includegraphics[scale=0.438]{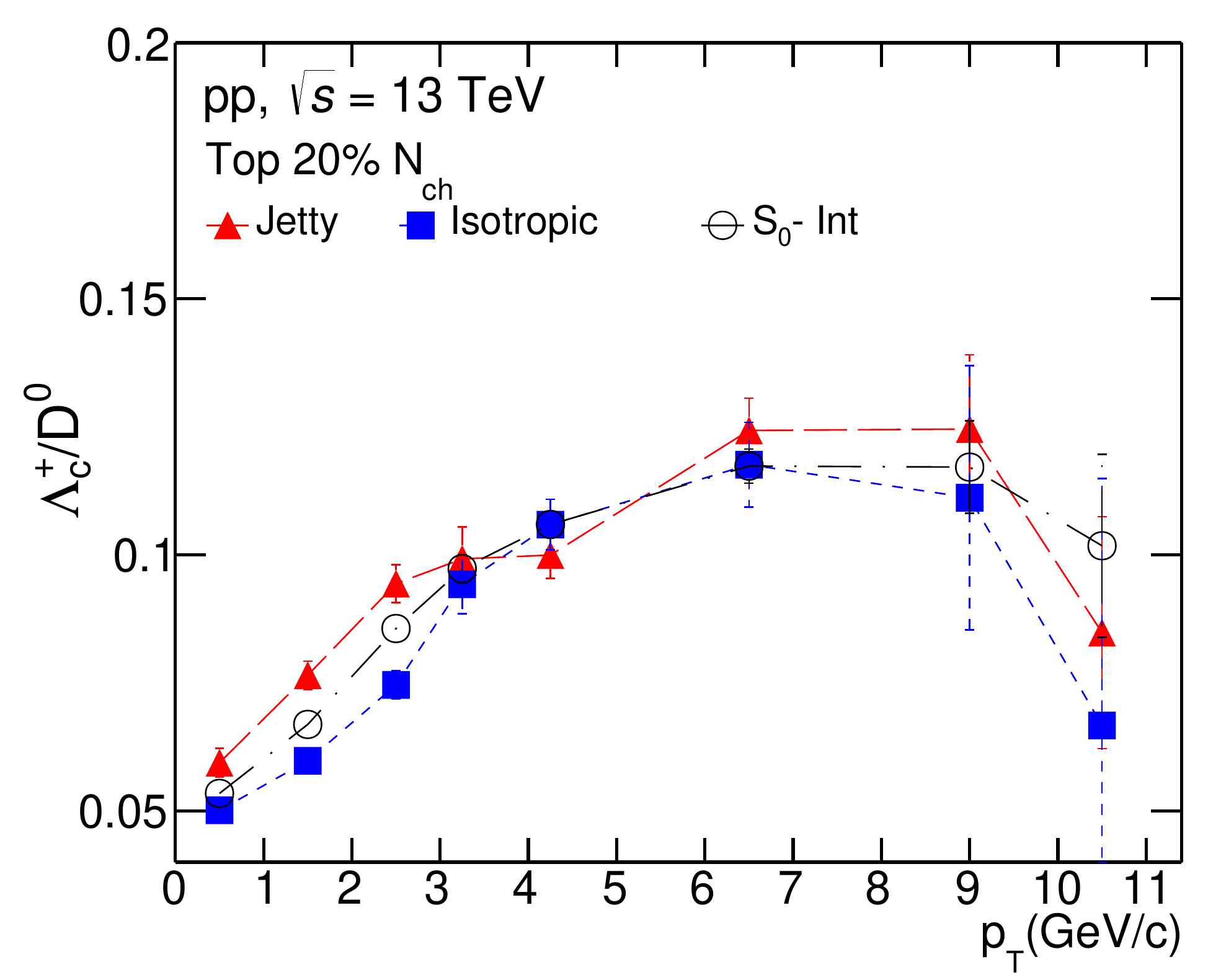}
\caption {(Color online) $p_{\rm{T}}$-differential particle ratio of  $\Lambda_{c}^{+}$ to $\rm D^{0}$, for minimum bias and high-multiplicity (top 20\%) pp collisions in isotropic (blue squares), jetty (red triangles) and spherocity integrated (open circles) events using PYTHIA8.}
\label{fig5:lambda_to_dzero}  
\ec
\end{figure}

A significant enhancement of baryon-to-meson ratios for light hadrons has been observed in central heavy-ion collisions compared to pp collisions in the intermediate $p_{\rm{T}}$ region~\cite{Abelev:2013xaa}. The enhancement can be explained by coalescence model through hadronize-combination of constituent quarks~\cite{Oh:2009zj,Greco:2003mm,Minissale:2015zwa}. Recently, ALICE and LHCb have observed enhancement of charmed baryon-to-meson ratio which indicates charm quarks may hadronize through coalescence as well. Although, minimum bias pp collisions do not show significant enhancement of baryon-to-meson ratio in the intermediate $p_{\rm{T}}$ region~\cite{Acharya:2020lrg,Aaij:2018iyy}, in this paper we have tried to unfold the possibility of such effects in high-multiplicity events in different event shapes. Indication of such enhancement would be sensitive to thermalization effect in pp collisions~\cite{Sarkar:2019nua}. The relative abundance of baryons and mesons can shed light on the process of fragmentation - a non-perturbative process. Formation of jets of partons into high transverse momentum hadrons is described by fragmentation function which incorporate how partons from jet combine with quarks and antiquarks from the vacuum to form hadrons. Because of MPIs, jet-partons in pp collisions can combine with quarks and antiquarks produced from MPIs to form hadrons via string fragmentation. Since the momenta of quarks and antiquarks from secondary MPIs are smaller than those of partons from jets, these hadrons have momenta lower than independent fragmentation of jet partons and that is what we observe from Fig.~\ref{fig5:lambda_to_dzero}. The  $p_{\rm T}$-differential $\Lambda^{+}_{c}/\rm D^{0}$ ratio for jetty events is higher compared to isotropic events in minimum bias sample.  One interesting observation from Fig.~\ref{fig5:lambda_to_dzero} is that the behaviour of $\Lambda^{+}_{c}/\rm D^{0}$ $p_{\rm T}$-differential ratio for all event topologies follow heavy-ion-like trend i.e.  enhancement of baryon-to-meson ratio in the intermediate $p_{\rm{T}}$ region followed by a decreasing behaviour. Although, the minimum bias samples show a clear event topology
dependence, the top 20\% high-multiplicity pp events are driven by the final state multiplicity without a distinction of event
types.

\begin{figure}[!ht]
\bc
\includegraphics[scale=0.438]{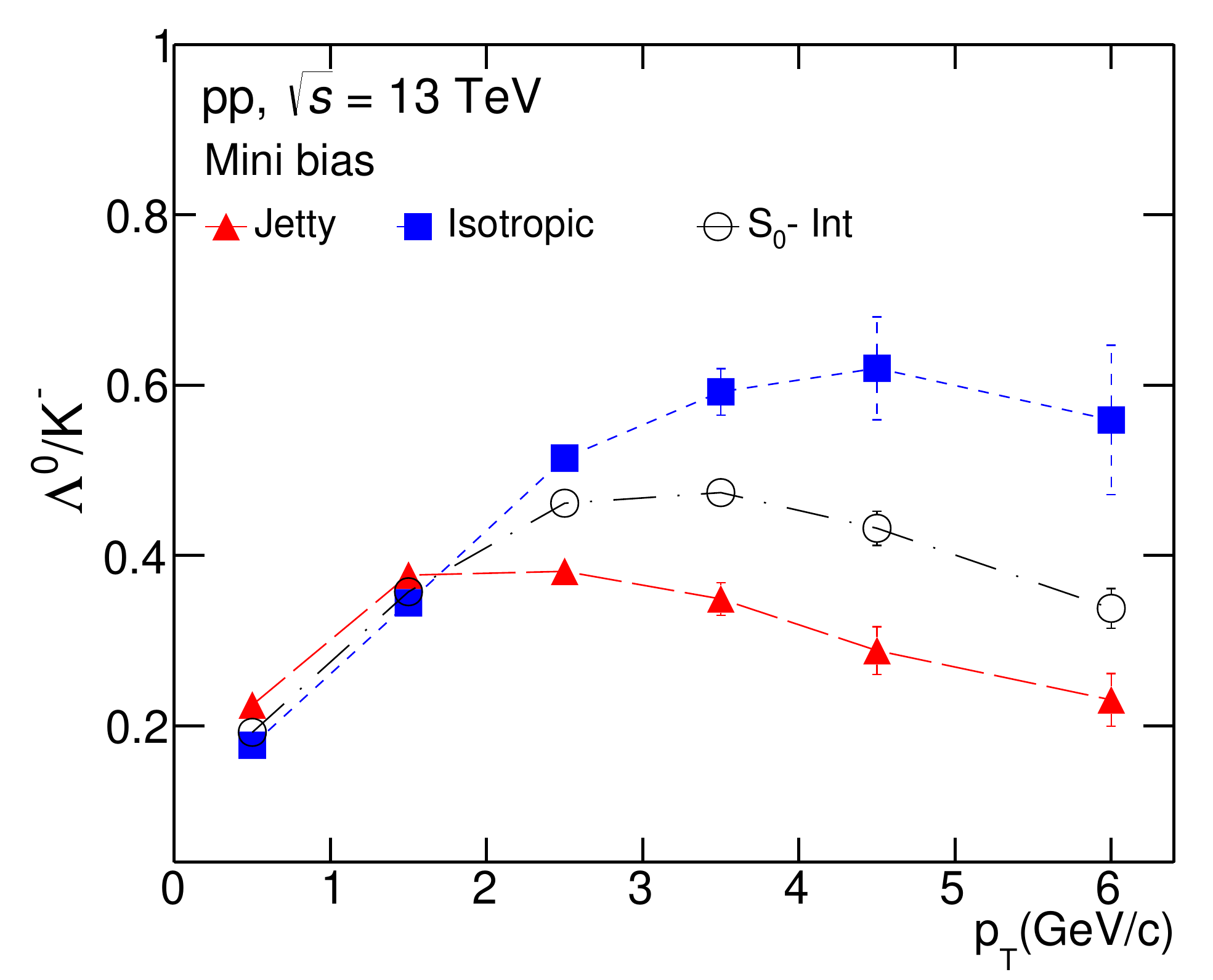}
\includegraphics[scale=0.438]{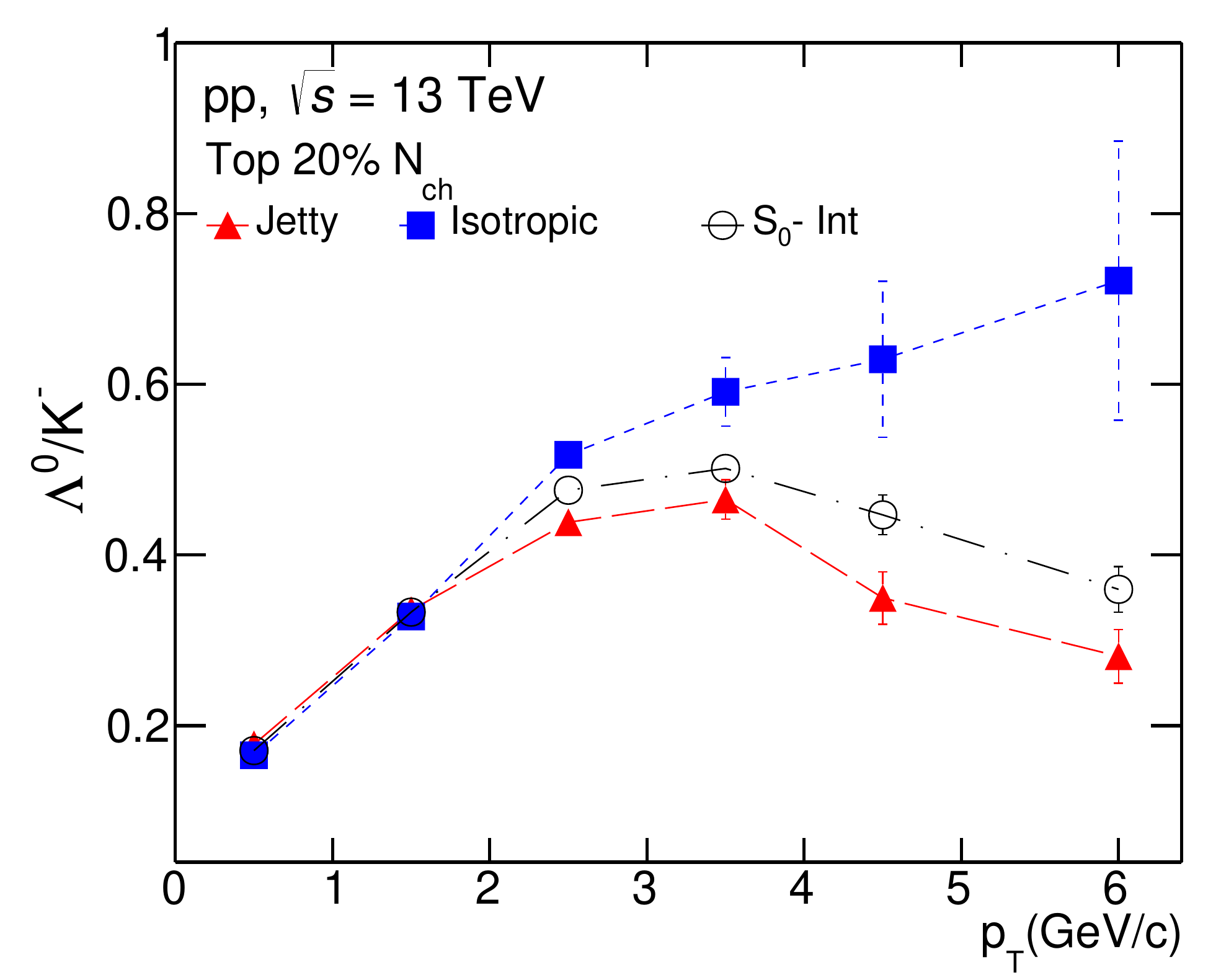}
\caption {(Color online) $p_{\rm{T}}$-differential particle ratio of  $\Lambda^{0}$ to $\rm K^{-}$, for minimum biased and high-multiplicity (top 20\%) pp collisions in isotropic (blue squares), jetty (red triangles) and spherocity integrated (open circles) events using PYTHIA8.}
\label{fig5:lambda_to_K}  
\ec
\end{figure}

Contrary to heavy-flavors, when we study similar ratio in the light flavor sector ($\Lambda^{0}/K^{-}$), we observe a completely opposite trend with spherocity classes for $p_{\rm T}$ $>$ 4 GeV/$c$ (Fig.~\ref{fig5:lambda_to_K}): ratio is higher for isotropic samples as compared to jetty ones. This is because of the fact that the former are driven predominantly by hard collisions and can have maximum contributions from jettiness of the events in comparison with the contributions from hadronization. However, for light flavors, most of the contributions could be MPI dominant. Here, the $\Lambda$ enhancement is linked to the increased density of quarks and gluons, particularly the strange quarks (s) from MPI and CR in the final state. 

For the  heavy flavor versus light flavor behaviour of the baryon over meson ratio, heavier particles will have a larger boost which will be
reflected in the baryon-to-meson ratios. 
Therefore, one can see the shift of peak of the ratio to higher $p_{\rm{T}}$ for heavy-flavors. For the top 20\% high-multiplicity pp
events, as seen from Figs.~\ref{fig5:lambda_to_dzero} and~\ref{fig5:lambda_to_K}, event topology has no effect on heavy-flavor sector, whereas in case of the light-flavors, we do observe a clear dependence of the discussed ratio on different event types.

 \section{Summary}
 \label{sum}
This paper focuses on the production of heavy-flavor hadrons --  of J/$\psi$, $\rm D^{0}$ and $\Lambda_{c}^{+}$ in pp collisions at $\sqrt{s}$ = 13 TeV using 4C tuned PYTHIA8 event generator at mid-rapidity. In addition, an event shape observable called spherocity is used for the first time in the heavy-flavor sector as a differentiator of integrated events into jetty and isotropic to have a better understanding of production dynamics of the heavy-flavor hadrons. 
Important findings from this study are summarized below:
 
 \begin{enumerate}
\item We see a clear dependence of the spherocity distribution with charged-particle multiplicity.  The spherocity distribution is increasingly skewed with the increase in charged-particle multiplicity.\\

\item A clear spherocity dependence of heavy-flavor $p_{\rm T}$-spectra, integrated yield, $\langle p_{\rm{T}}\rangle$ and particle ratios is observed in both minimum bias and high-multiplicity pp collisions.\\
 
 \item The crossing point of the ratios of $p_{\rm T}$-spectra from jetty and isotropic events to the spherocity-integrated ones shifts to lower $p_{\rm T}$ with the increase in charged-particle multiplicity. This indicates that spherocity differentiates events (jetty versus isotropic) more accurately in high-multiplicity pp collisions keeping a small gap in the multiplicity of heavy-flavor hadrons.\\
 
 \item Relative yield and relative $\langle p_{\rm{T}}\rangle$ are found to be increasing with the increase in charged-particle multiplicity and they are higher for jetty events compared to isotropic ones. These results suggest that spherocity acts as a nice tool to differentiate events dominated by soft versus hard particle production processes.

\item The spherocity dependence of relative yields and relative $\langle p_{\rm{T}}\rangle$ for $\rm D^{0}$ and $\Lambda_{c}^{+}$ show a similar trend while for J/$\psi$ the difference from jetty to isotropic events is found to be lesser. This novel observation hints to different production dynamics of open charm compared to charmonia and the MPIs with color reconnection mechanism plays a major role for such a behavior in PYTHIA8. \\

\item The $\Lambda^{+}_{c}/\rm D^{0}$ ratio in jetty events is found to be higher compared to the isotropic events while an opposite trend for $\Lambda^{0}/K^{-}$ ratio is observed for the minimum bias sample. This is an interesting observation as spherocity dependence of particle ratios show a completely different behaviour for heavy flavor compared to light flavor sector. This clearly indicates to a MPI dominant contribution for $\Lambda^{0}/K^{-}$ while the $\Lambda^{+}_{c}/\rm D^{0}$ ratio is driven predominantly by hard collisions and can have maximum contributions from jets.\\

\end{enumerate}
 
 A multi-differential study taking event topology and multiplicity is necessary in small systems at LHC energies when looking into  the observation of heavy-ion like features in  high-multiplicity pp collisions. The LHC experiments have planned for a dedicated high-multiplicity triggered events and the associated detector upgrades that will provide a proper platform in this direction. Study of heavy-flavor production will play an important role for the test of the pQCD, as they are produced early in time and witness the complete spacetime evolution of the system. However, the present limitations in terms of proper identification of secondary vertices, efficiency at low-$p_{\rm T}$ and dealing with signal to background ratio will be overcome to a greater extent with the detector upgrades. It is worth mentioning here that ALICE ITS3 planned for installation in LHC Long Shutdown3 (LS3), will have a novel vertex detector consisting
of curved wafer-scale ultra-thin silicon sensors arranged in perfectly cylindrical layers. This will feature an unprecedented low material budget of 0.05\% X0 per layer, with the innermost layer positioned at
only 18 $mm$ radial distance from the interaction point \cite{LS3,Future}. This will help with higher efficiency of detection of heavy-flavor particles, opening up a new domain of pQCD studies. The present study will be more exciting to carry out in experimental data in the upcoming LHC Run-3 and Run-4.
  
\section{Acknowledgement} 
 This research work has been carried out under the financial supports from DAE-BRNS, Government of India, Project No. 58/14/29/2019-BRNS of Raghunath Sahoo. S.T. acknowledges the support from the postdoctoral fellowship of DGAPA UNAM.


\setcounter{secnumdepth}{0}
\appendix
\section{Appendix}
\label{appendix}
\begin{figure*}[!ht]
\bc
 \includegraphics[scale=0.44]{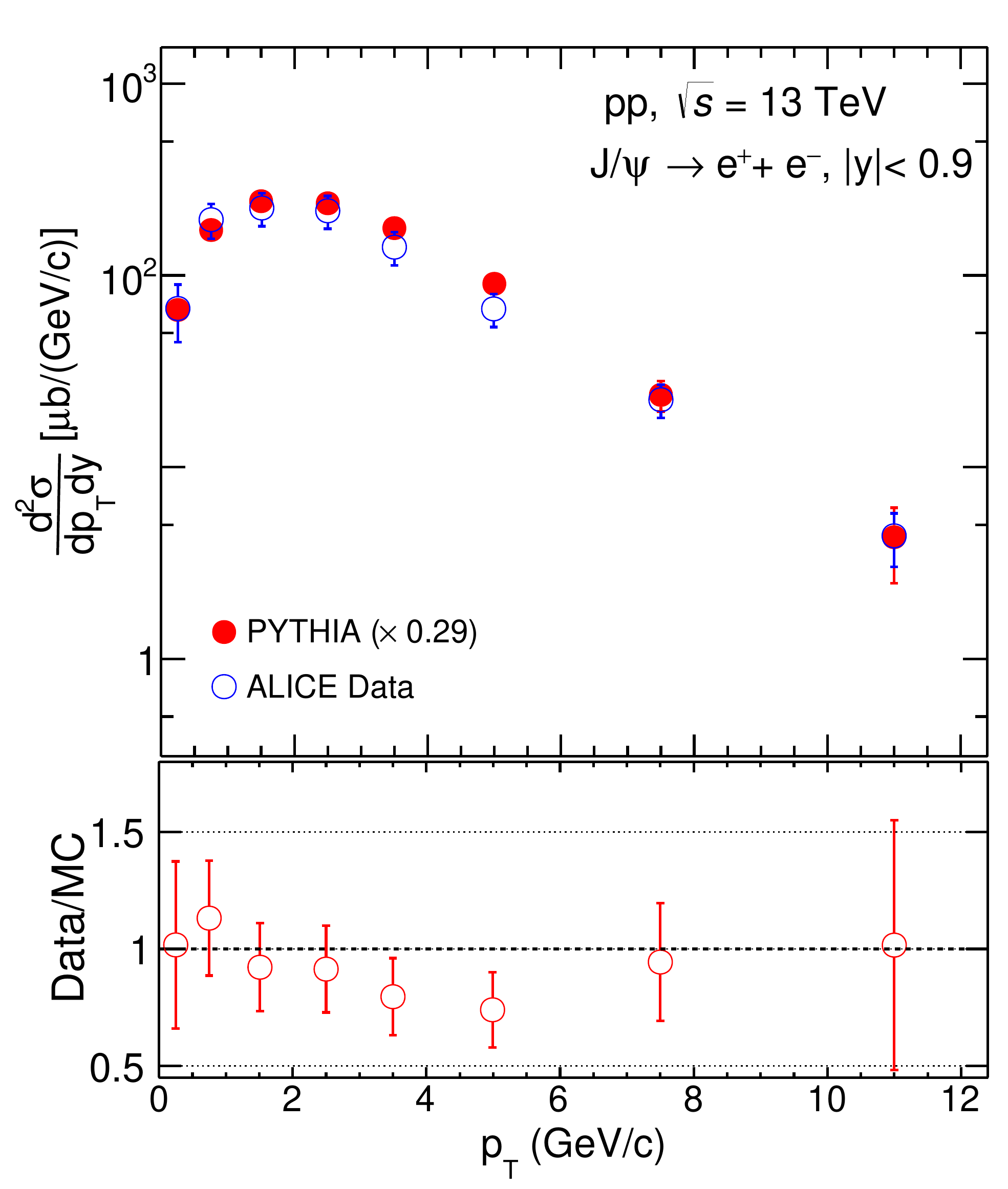}
  \includegraphics[scale=0.44]{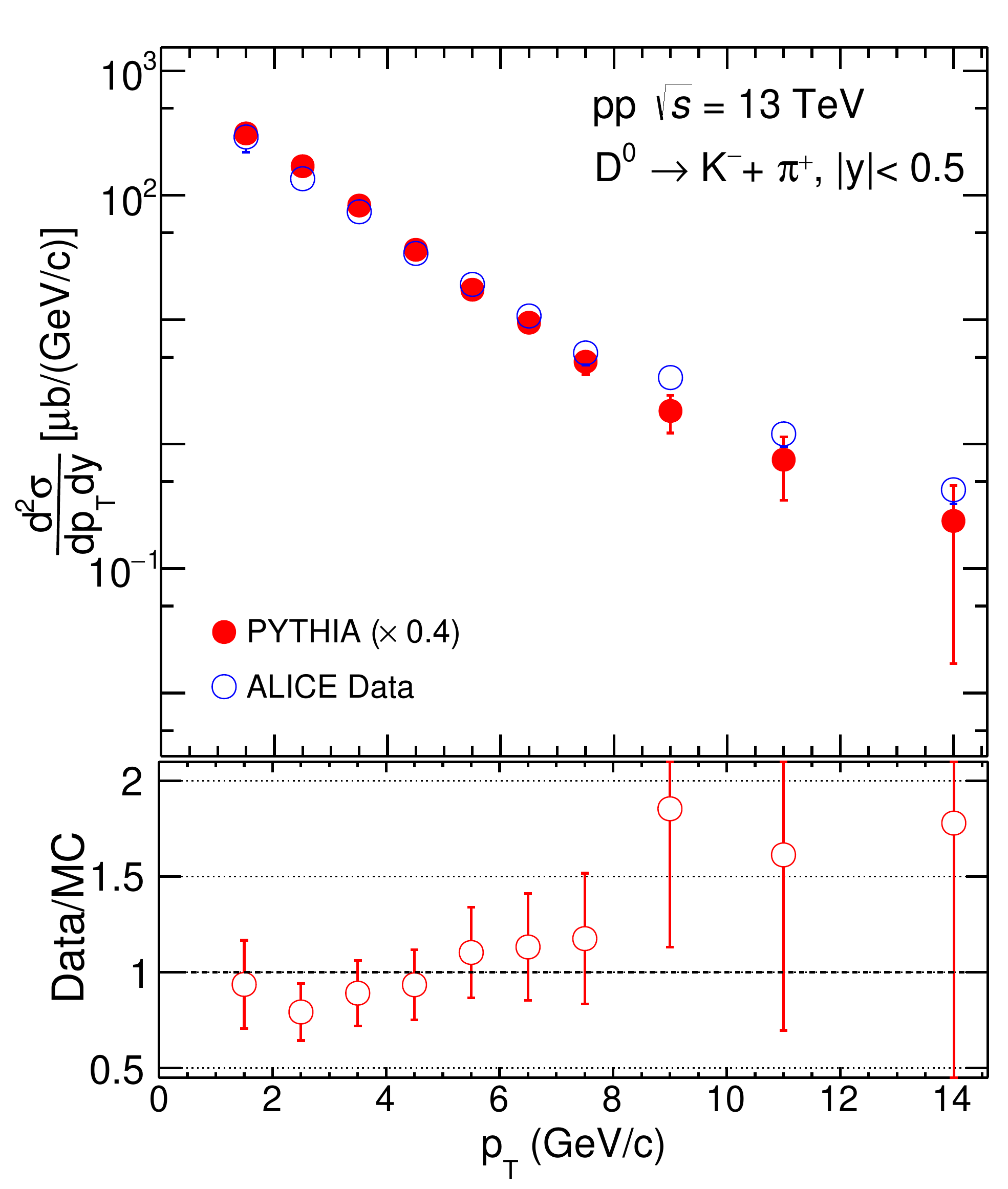}
 \caption  {(Color online) Top panel shows the comparison of ALICE data~\cite{Lofnes,Hamon:2018zqs} and PYTHIA8 of J/$\psi$ (left) and $\rm D^{0}$ (right) production cross-section as a function of transverse momentum ($p_{\rm{T}}$) for pp collisions at $\sqrt{s}$ = 13 TeV. The open blue circles are ALICE data and solid red circles represent PYTHIA8 results. The quadratic sum of statistical and systematic uncertainties of ALICE data are presented in a single error bar. Bottom panels show the ratio between ALICE data and PYTHIA8, and the error bars are estimated using standard error propagation formula.}
 \label{fig:jpsinD0}  
 \ec
 \end{figure*}

To check the compatibility of PYTHIA8 with the experimental data, we have compared the production cross-sections of J/$\psi$ and $\rm D^{0}$ between experimental data from ALICE and PYTHIA8 in the same kinematic range. Left (Right) panel of Fig.~\ref{fig:jpsinD0} shows the  comparison  of  J/$\psi$ ($\rm D^{0}$) production cross-section in pp collisions as a function of $p_{\rm{T}}$, respectively for minimum bias events. The open symbols represent the data obtained from ALICE experiment~\cite{Lofnes}(\cite{Hamon:2018zqs}) and the solid circles show the results from PYTHIA8 event generator in pp collisions at $\sqrt{s}$ = 13 TeV. In order to see how well the spectral shapes obtained from PYTHIA8 simulation match with the experimental data, we have used some arbitrary multipliers. Within uncertainties, PYTHIA8 seems to reproduce similar spectral shapes as from experimental data for both J/$\psi$ and $\rm D^{0}$.

 \end{document}